\documentclass[doublecol]{epl2_arxiv} 
\usepackage[utf8]{inputenc}
\usepackage[T1]{fontenc}
\usepackage{xcolor}
\usepackage{amsmath}
\usepackage{amssymb}

\definecolor{darkblue}{rgb}{0,0,0.6}
\definecolor{darkred}{rgb}{0.6,0,0}
\usepackage[colorlinks=true,urlcolor=darkblue,citecolor=darkblue,linkcolor=darkred,hyperfootnotes=false]{hyperref}

\newcommand{\ind}[1]{_{\mathrm{#1}}}

\newcommand{\bR}{\mathbb{R}}

\newcommand{\dd}{\mathrm{d}}

\begin{document}

\title{Phase diagram of a bulk 1d lattice Coulomb gas}

\author{V. Démery\inst{1} \and R. Monsarrat\inst{1} \and D. S. Dean\inst{2} \and R. Podgornik\inst{3,4}}
\shortauthor{V. Démery \etal}

\institute{            
  \inst{1} PCT, UMR Gulliver 7083, ESPCI and CNRS, 10 rue Vauquelin, 75005 Paris, France.\\
  \inst{2} Universit\'e de  Bordeaux and CNRS, Laboratoire Ondes et
Mati\`ere d'Aquitaine (LOMA), UMR 5798, F-33400 Talence, France.\\
  \inst{3} Department of Physics, Faculty of Mathematics and Physics, University of Ljubljana, SI-1000 Ljubljana, Slovenia.\\
  \inst{4} Department of Theoretical Physics, J. Stefan Institute, SI-1000 Ljubljana, Slovenia.
}

\abstract{
The exact solution, via transfer matrix, of the simple one dimensional lattice Coulomb gas (1d LCG) model can reproduce peculiar features of ionic liquid capacitors, such as overscreening, layering, and camel- and bell-shaped capacitance curves. Using the same transfer matrix method, we now compute the bulk properties of the 1d LCG in the constant voltage ensemble.  We unveil a phase diagram with rich structure exhibiting a low density disordered and high density ordered phases, separated by a first order phase transition at low temperature; the solid state at full packing can be ordered or not, depending on the temperature. This phase diagram, which is strikingly similar to its three dimensional counterpart, also sheds light on the behaviour of the confined system.}

\maketitle


Ionic liquids have intriguing properties when used in capacitors. These include  
self-organization in alternating positively and negatively charged layers close to a charged surface~\cite{Mezger2008,Hayes2011,Perkin2011,Perkin2012},
overscreening of the surface charge~\cite{Fedorov2014} and camel- and bell-shaped capacitance curves depending on temperature~\cite{Fedorov2008}.
The first attempt to explain these features was based on a model with electrostatic interactions and steric repulsion treated at the mean-field level. This treatment successfully predicted the camel and bell-shaped capacitance curves~\cite{Borukhov2000,Kornyshev2007}, however additional terms are required to explain the layering~\cite{Bazant2011}. 
Later, a 1d lattice model was solved exactly, showing that electrostatic interaction and steric repulsion  alone can qualitatively reproduce  the observed features~\cite{Demery2012,Demery2012a}.
A continuum version of the one dimensional model, modeling the ions as charged rods, showed a better qualitative agreement with experimental electrode separation-pressure curves~\cite{Lee2014}. However, in this case, an exact solution was not found, and a numerical computation is only possible for small systems. All these studies of one dimensional systems focused on the capacitor behaviour while their bulk properties remained unexamined. 
We now extend this analysis to the bulk system with the rationale that the understanding of bulk properties may potentially provide valuable insight into the behaviour of small systems.

The basic ingredients of ionic liquids, electrostatic interaction and steric repulsion, are incorporated in the restricted primitive model (charged hard spheres), the phase diagram of which has been extensively studied and is now well established~\cite{Stell1992,Orkoulas1994,Vega1996,Stell1999,Bresme2000,Luijten2002,Vega2003,Panagiotopoulos2005}. 
It features a phase coexistence between a low density gas and a high density ordered liquid at low temperature, a Néel line of second order transitions at larger temperatures, and a solid phase at high density.
The lattice version of this model (LRPM) implements the steric exclusion by allowing at most one ion per site. 
Its phase diagram has the same structure as its continuous counterpart, but also exhibits significant differences stemming from the fact that the lattice facilitates organization~\cite{Walker1983,Panagiotopoulos1999,Dickman1999,Brognara2002, Kobelev2002, Ciach2003, Panagiotopoulos2005,Diehl2005,Ren2006,Bartsch2015}.
The bulk properties of RPM and LRPM are thus well understood, but the link with the peculiar behaviour of an ionic liquid capacitor remains to be made. 

Here, we use the exact transfer matrix solution of the one dimensional LRPM (or one dimensional lattice Coulomb gas) developed in Refs.~\cite{Demery2012,Demery2012a} in the context of a fluid capacitor to compute its bulk phase diagram. We show that, remarkably, this simple model produces a rich phase diagram showing striking similarities to its 3d counterpart. 
Specifically, we find that the charge density oscillations observed close to a charged surface correspond to an ordered bulk phase, and the abrupt change of the capacitance at the point of zero charge~\cite{Demery2012a} of an ionic liquid capacitor corresponds to a first order bulk phase transition.

\section{Model}\label{}

\begin{figure}
\begin{center}
\includegraphics[scale=1.25]{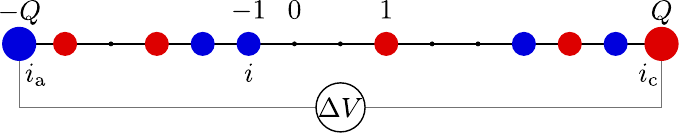}
\end{center}
\caption{(Colour on-line) Notations for the one-dimensional lattice Coulomb gas: sites can either be empty, host a cation (red) or an anion (blue). Electrodes bear charges $-Q$ and $Q$.}
\label{fig:scheme}
\end{figure}

We begin by defining the one-dimensional lattice Coulomb gas model, shown in Fig.~\ref{fig:scheme}.
The ions are confined between an anode at $i\ind{a}=\lceil-(N+1)/2\rceil$ and a cathode at $i\ind{c}=\lceil(N+1)/2\rceil$, $N$ sites are thus available.
Each site $i\in (i\ind{a},i\ind{c})$ can be empty ($s_i=0$), host a cation ($s_i=1$) or an anion ($s_i=-1$).
In the constant charge ensemble, the cathode and anode bear a charge $Q\in\mathbb{R}_+$ and $-Q$, respectively.
The grand partition function can be written as
\begin{equation}
\Xi = \sum_{(s_i)} \delta_{0,\sum_i s_i} \mu^{\sum_i|s_i|}\exp \left(\frac{1}{4T}\sum_{i,j}|i-j|s_is_j \right),
\end{equation}
where the index $i$ runs over $(i\ind{a},i\ind{c})$, except in the argument of the exponential where the sum includes the electrodes, with $s_{i\ind{a}}=-Q$ and $s_{i\ind{c}}=Q$. 
In what follows, we use a unit charge, unit lattice spacing, and unit Boltzman constant: the only free parameters are thus the fugacity $\mu$ and the temperature $T$. 
Electroneutrality is enforced via the Kronecker $\delta$.
If the electrodes can exchange charges to maintain a potential difference $\Delta V$ between them, the relevant ensemble is the \emph{constant voltage ensemble}. In this ensemble, the grand partition function is given by~\cite{Demery2012a}
\begin{equation}\label{eq:pf_cv}
\hat\Xi_{\Delta V} = \int_{-\infty}^\infty \exp(Q\Delta V)\Xi_Q\dd Q.
\end{equation}

\section{Transfer matrix solution}\label{}

The model defined above is exactly solvable using a Hubbard-Stratonovitch transformation, which allows the rewriting of the partition function \emph{via} the transfer matrix formalism~\cite{Demery2012,Demery2012a}. The constant charge partition function thus reads
\begin{equation}
\Xi = \langle \psi\ind{b}|K^N|\psi\ind{b} \rangle,
\end{equation}
where the vector $\psi\ind{b}$ is a function and the transfer matrix $K$ is a linear integral operator defined on functions in $\bR$:
\begin{align}
\psi_{\mathrm{b},k} & = \exp \left(-\frac{k^2}{4T} \right)\delta(k-Q) \label{eq:bv}\\
K_{kl} & =\exp \left(-\frac{k^2+l^2}{4T} \right) \label{eq:tm}\\
& \qquad \times (\delta(k-l)+\mu[\delta(k-l+1)+\delta(k-l-1)]). \nonumber
\end{align}
The moments of the density can be computed using the cation ($+$) and anion ($-$) density operators defined as
\begin{equation}\label{eq:density_op}
R_{\pm,kl} = \mu\exp \left(-\frac{k^2+l^2}{4T} \right)\delta(k-l\pm 1).
\end{equation}
The charge and occupancy operators are given by $R=R_+-R_-$ and $R\ind{o}=R_++R_-$, respectively.
In the following, we will be interested in the average density and the charge-charge correlation function, which are given by
\begin{align}
\langle |s_i| \rangle & = \Xi^{-1} \langle \psi\ind{b}|K^{i-i\ind{c}-1}R\ind{o}K^{i\ind{a}-i-1}|\psi\ind{b} \rangle,\\
\langle s_i s_j \rangle & = \Xi^{-1} \langle \psi\ind{b}|K^{i-i\ind{c}-1}RK^{j-i-1}RK^{i\ind{a}-j-1}|\psi\ind{b} \rangle.
\end{align}

In the thermodynamic limit, $N\to\infty$, the maximal eigenvalue $\lambda^*$ of the transfer matrix and its associated eigenvector $\psi^*$ dominate, we can thus write $K^N\sim {\lambda^*}^N|\psi^*\rangle\langle\psi^*|$ and it follows that
\begin{align}
\Xi & \sim {\lambda^*}^N |\langle \psi\ind{b}|\psi^*\rangle|^2,\label{eq:thermo_pf}\\
\langle |s_i| \rangle & \to {\lambda^*}^{-1} \langle \psi^*|R\ind{o}|\psi^* \rangle=\rho,\label{eq:thermo_dens}\\
\langle s_i s_j \rangle & \to {\lambda^*}^{-(j-i+1)} \langle \psi^*|RK^{j-i-1}R|\psi^* \rangle = \rho c_{j-i},\label{eq:thermo_correl}
\end{align}
where we have defined the average occupancy $\rho$ and charge correlation function $c_i$.

The transfer matrix $K$ and density operators $R_\pm$ (Eqs.~(\ref{eq:tm}-\ref{eq:density_op})) map the subspaces $\mathcal{E}_\theta=\{\psi_k, k\in\mathbb{Z}+\theta\}$, where $\theta\in[0,1)$, onto themselves. 
Hence, for each $\theta$, a discrete (but still infinite) matrix can be associated with the transfer matrix restricted to the subspace $\mathcal{E}_\theta$, $K_\theta=\left.K\right|_{\mathcal{E}_\theta}$.
For constant charges $\pm Q$ on the electrodes, the boundary vector is given by Eq.~(\ref{eq:bv}), and only the subspace $\mathcal{E}_{Q-\lfloor Q\rfloor}$ is explored. 
This means that the eigenvalue (and corresponding eigenvector) entering Eqs.~(\ref{eq:thermo_pf}-\ref{eq:thermo_correl}) is actually the maximal eigenvalue of $K_{Q-\lfloor Q\rfloor}$, which we denote $\lambda^*_{Q-\lfloor Q\rfloor}$ (it is positive from the Perron-Frobenius theorem).
This dependence of the bulk behaviour of the system on the boundary conditions of a finite system is related to the $\theta$-vacua encountered in quantum field theory~\cite{Aizenmann1981}.

On the other hand, in the constant voltage ensemble, the partition function (Eq.~(\ref{eq:pf_cv})) obviously involves all the possible boundary charges.
As a consequence, the maximal eigenvalue $\lambda^*$ which enters Eqs.~(\ref{eq:thermo_pf}-\ref{eq:thermo_correl}) is the maximal eigenvalue of the full transfer matrix $K$. 
The charges $Q\in\mathbb{Z}+\theta$ such that $\lambda^*_\theta=\lambda^*$ are said to be 
``selected'' (we also say that a value of $\theta$ is selected). 

\begin{figure}
\begin{center}
\includegraphics[]{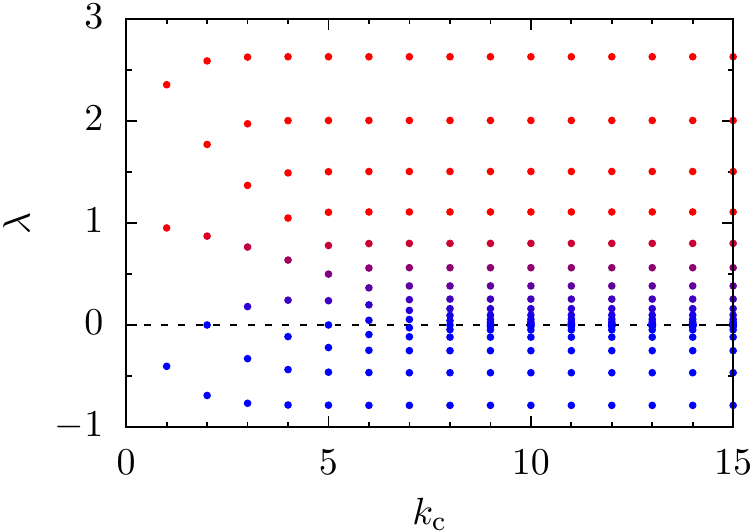}
\end{center}
\caption{(Colour on-line) Spectrum of the transfer matrix $K_{\theta=0}$ as a function of the numerical cutoff $k\ind{c}$ for $T=10$, $\mu=1$. The $2k_c+1$ eigenvalues corresponding to each value of $k_c$ are plotted along the corresponding $y$-axis and are color coded, blue denoting the smallest and red the largest. One clearly sees the relative insensitivity of $\lambda^*$ to $k_c$ for $k_c >5$.}
\label{fig:spec_modes}
\end{figure}

For each $\theta$, the transfer matrix $K_\theta$ has infinite dimension, so that a cut-off $k\ind{c}$ has to be introduced on its indices in order to diagonalise it numerically. 
The spectrum of $K_{\theta=0}$ for $T=10$ and $\mu=1$ is shown in Fig.~\ref{fig:spec_modes}; increasing the number of modes beyond $k\ind{c}\simeq 5$ increases the density of eigenvalues around 0, but does not change the maximal (and minimal) eigenvalues, neither the spectral gap between the first and second eigenvalues. 
This rapid convergence with the cutoff is due to the rapid decay of the transfer matrix coefficients with the indices (see Eq.~(\ref{eq:tm})).
The accuracy of the evaluation is better than the standard precision of Python, $10^{-15}$, for the cut-off $k\ind{c}=25$ used throughout the manuscript.

\begin{figure}
\begin{center}
\includegraphics[]{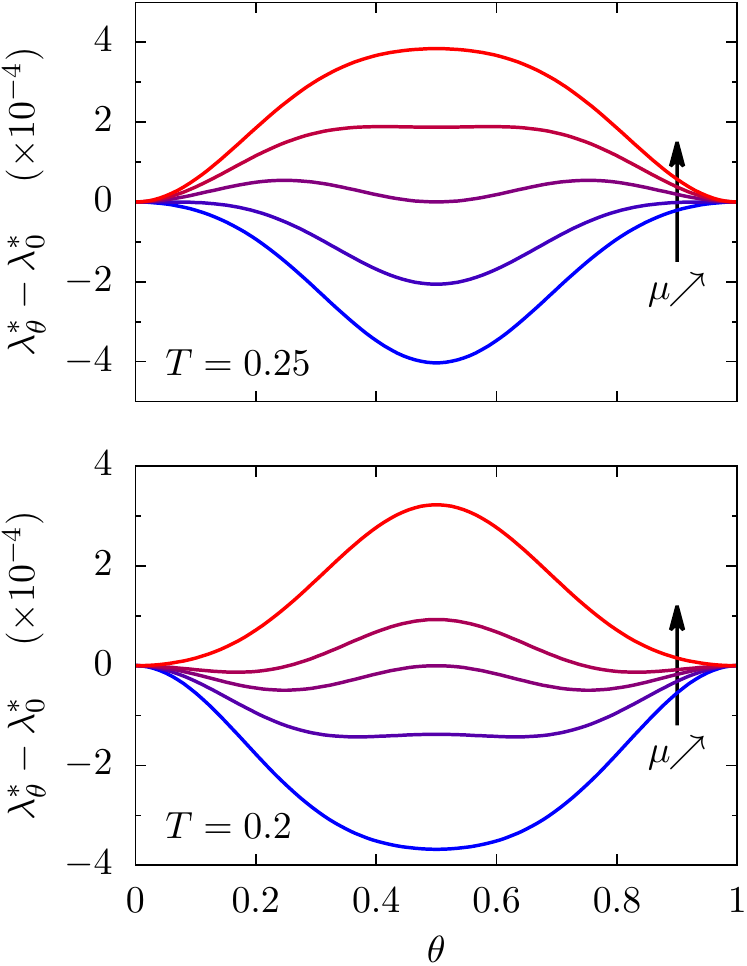}
\end{center}
\caption{(Colour on-line) 
Dependence of the maximal eigenvalue $\lambda_\theta^*$ on $\theta$ at $T=0.25$ (\emph{top}) and $T=0.2$ (\emph{bottom}) around a value of the fugacity $\mu$ where the selected $\theta$ goes from $0$ to $1/2$ (top: $\mu\in\{1.122,1.123,1.12404664,1.125,1.126\}$; bottom: $\mu\in\{1.324,1.325,1.32559787,1.326,1.327\}$).
For $T=0.25$ the selected value of $\theta$ goes continuously from $0$ to $1/2$ while the transition is sharp for $T=0.2$.
}
\label{fig:lambdamax_theta}
\end{figure}

The relations $\lambda^*_\theta=\lambda^*_{-\theta}$ and $\lambda^*_\theta=\lambda^*_{1+\theta}$ mean that $\theta =0$ and $1/2$ are stationary points of $\lambda^*_\theta$ with respect to $\theta$. 
Numerically we find that integer ($\theta=0$) or half-integer ($\theta=1/2$) boundary charges correspond to the maximal value of $\lambda^*_\theta$ almost everywhere.
Curves of $\lambda^*_\theta$ are shown for values of the fugacity close to a transition on Fig.~\ref{fig:lambdamax_theta}. Depending on the shape of these curves, the selected $\theta$ can either change continuously (\emph{e.g.}, for $T=0.25$) or discontinuously (\emph{e.g.}, for $T=0.2$); the latter corresponds, as we shall see, to a first order phase transition.
On the contrary, no such transition exists in the constant charge ensemble. 
We are interested in the bulk phase transitions that may take place in the system, so that we restrict our study to the constant voltage ensemble; 
in the thermodynamic limit, bulk averages do not depend on the imposed voltage (see Eqs.~(\ref{eq:thermo_dens}-\ref{eq:thermo_correl})).
Note that the occurrence of the free energy minima at $\theta=0$ for low density systems and $\theta=1/2$ for high density systems can be partially explained by mapping the model here onto a discrete interface model~\cite{Demery2012a}.

\section{Density}\label{}

\begin{figure}
\begin{center}
\includegraphics[scale=1.]{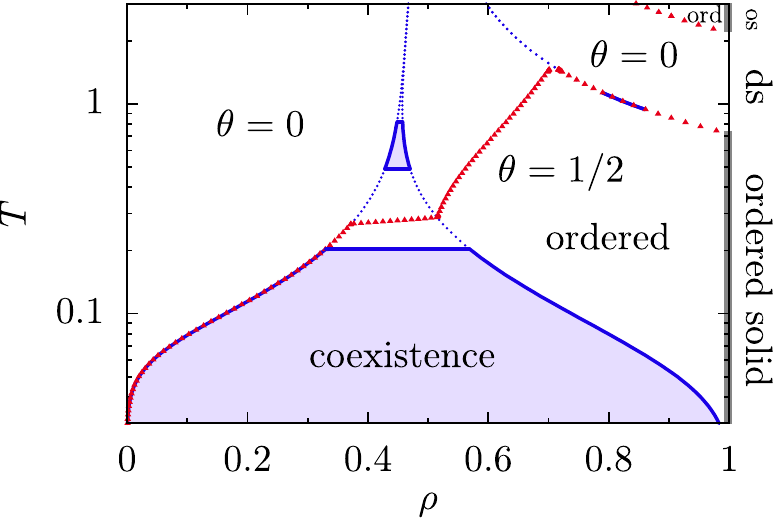}
\end{center}
\caption{(Colour on-line) Phase diagram in the density-temperature plane. 
Blue dotted lines delineate regions of integer ($\theta=0$) and half-integer ($\theta=1/2$) selected charges.
Solid blue lines delineate the coexistence regions.
Red triangles denote the onset of charge oscillation, i.e., ordering. 
The solid phase at $\rho=1$ is ordered for $\theta=1/2$ (gray lines, os) and disordered for $\theta=0$ (ds).
}
\label{fig:phase_diag}
\end{figure}

When the selected charges switch discontinuously between integers and half-integers, as the chemical potential increases, the density changes discontinuously (from $\rho\ind{a}$ to $\rho\ind{b}$). 
At imposed density, this discontinuity becomes a coexistence region for $\rho\in[\rho\ind{a},\rho\ind{b}]$.
Coexistence should be understood in the statistical sense: the statistical state of the system is a combination of low and high density states. In this region, the maximal eigenvalue $\lambda^*$ of the transfer matrix and the pressure $p=T\log(\lambda^*)$ do not depend on the density.
The density-temperature phase diagram, shown on Fig.~\ref{fig:phase_diag}, displays three coexistence regions. 
The main one, at low temperature, is reminiscent of the one present in the RPM~\cite{Bresme2000}. 
The one at higher temperature along the same transition line and the one around $T\simeq 1$, $\rho\simeq 0.8$ are more exotic.

\section{Spatial organization}\label{}

\begin{figure}
\begin{center}
\includegraphics[]{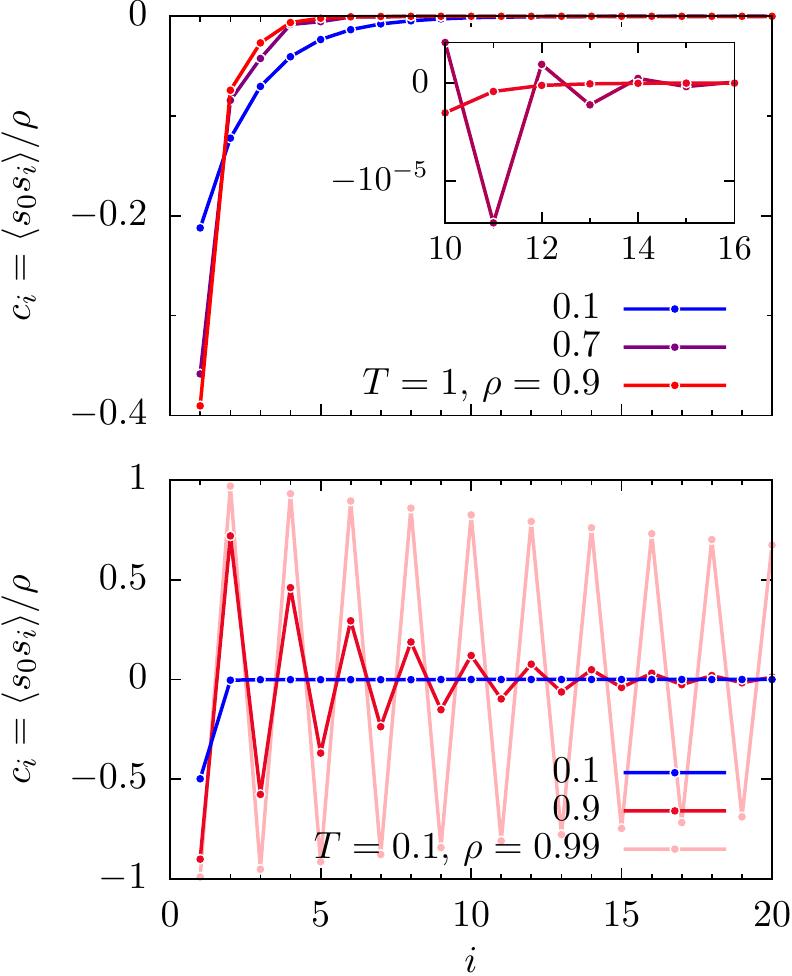}
\end{center}
\caption{(Colour on-line) Charge correlation function as a function of the distance. \emph{Top:} $T=1$ and $\rho\in \{0.1,0.7,0.9\}$; \emph{inset:} focus on $\rho\in\{0.7,0.9\}$ at large distance. \emph{Bottom:} $T=0.1$ and $\rho\in\{0.1,0.9,0.99\}$.
}
\label{fig:correls}
\end{figure}

Spatial organization of the charge is encoded in the charge correlation function $c_i$.
The correlation function is shown on Fig.~\ref{fig:correls} for several densities and temperatures;
it may decay monotonically or with oscillations. We focus on the long distance decay to understand these two behaviours.

The long distance behaviour of the correlation function can be investigated by keeping only the two maximal eigenvalues of $K$ (in absolute value), $\lambda^*$ and $\lambda_2$, corresponding to the eigenvectors $\psi^*$ and $\psi_2$, respectively: 
\begin{equation}
c_i\underset{i\to\infty}{\sim} \frac{\langle \psi^*|R\left({\lambda^*}^{i-1}|\psi^*\rangle\langle\psi^*|+\lambda_2^{i-1}|\psi_2\rangle\langle\psi_2|\right)R|\psi^*\rangle}{{\lambda^*}^{i+1}\rho}.
\end{equation}
The system is neutral, $0=\langle s_0 \rangle=\langle\psi^*|R|\psi^*\rangle/\lambda^*$, and $R$ is antisymmetric, so that 
\begin{equation}\label{eq:correl_psi2}
c_i\underset{i\to\infty}{\sim}- \frac{\left| \langle\psi_2|R|\psi^*\rangle \right|^2}{{\lambda^*}^2\rho}\left(\frac{\lambda_2}{\lambda^*} \right)^{i-1}.
\end{equation}
Note that this relation breaks down if $\langle\psi_2|R|\psi^*\rangle=0$; in this case the largest eigenvalue such that the corresponding wavevector $\psi\ind{eig}$ satisfies $\langle\psi^*|R|\psi\ind{eig}\rangle\neq 0$ should be considered.
If $\lambda_2<0$ (i.e., $\lambda_2$ is the minimal eigenvalue of $K$) and $\langle\psi_2|R|\psi^*\rangle\neq 0$, the correlation function oscillates.
We find numerically that this last condition is never satisfied in the phase $\theta=0$, but is satisfied in the phase $\theta\neq 1/2$ at large enough density. 
The charge oscillations are seen on Fig.~\ref{fig:correls} for $T=1$ and $\rho=0.7$ ($\theta=1/2$ for this density, while $\theta=0$ for $\rho\in\{0.1,0.9\}$) and for $T=0.1$ and $\rho\in\{0.9,0.99\}$.
We denote the onset of charge oscillation, and thus organization, by red triangles on the phase diagram, Fig.~\ref{fig:phase_diag}.
The correlation length is given by $\xi = \log(|\lambda^*/\lambda_2|)^{-1}$, which appears to diverge as the density approaches $1$ (see Fig.~\ref{fig:correls} for $T=0.1$). This is investigated later using an expansion around $\rho=1$.

\section{Dense packing limit}

\begin{figure}
\begin{center}
\includegraphics[]{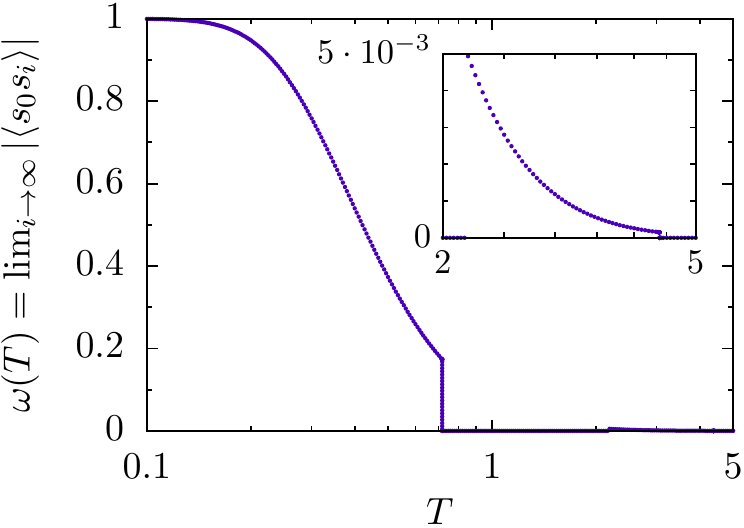}
\end{center}
\caption{(Colour on-line) Long-range charge order as a function of temperature for $\rho=1$. Inset is a zoom on the $T>1$ region, where a small long-range order exists.}
\label{fig:solidity}
\end{figure}

The dense packing limit $\rho\to 1$ corresponds to $\mu\to\infty$.
In this limit, some analytical results may be obtained from the properties of $\bar K=K/\mu$ and $\bar R_\pm=R_\pm/\mu$; the eigenvalues now refer to eigenvalues of $\bar K$.
First, we focus on the case of dense packing, i.e., $\rho=1$, where the system is in a solid state.
From the eigenvector $\psi^*$, we construct the vector $\psi_-$ with components $(\psi_-)_k=(-1)^{\lfloor k \rfloor}\psi^*_k$. From $\bar K_{kl}(-1)^{\lfloor l \rfloor} = - (-1)^{\lfloor k \rfloor}\bar K_{kl}$, we deduce that $\bar K |\psi_-\rangle=-\lambda^*|\psi_-\rangle$: $-\lambda^*$ is also an eigenvalue of $\bar K$. 
Hence $\lambda_2=-\lambda^*$ in Eq.~(\ref{eq:correl_psi2}), and we get
\begin{equation}
c_i\underset{i\to\infty}{\sim} \frac{\left| \langle\psi_-|\bar R|\psi^*\rangle \right|^2}{{\lambda^*}^2}(-1)^i.
\end{equation}
The parameter $\omega(T)={\lambda^*}^{-2}\left| \langle\psi_-|\bar R|\psi^*\rangle \right|^2$ characterises the long-range charge order. It depends only on temperature. 
If $\omega(T)>0$, the system is an ordered solid.
The long-range charge order is plotted in Fig.~\ref{fig:solidity}: it is zero for $\theta=0$ and non-zero for $\theta\neq 1/2$, although it decays rapidly with increasing temperature; the ordered and disordered solid regions are shown on the phase diagram, Fig.~\ref{fig:phase_diag}.
While ordered and disordered solid regions exist in the RPM~\cite{Vega1996,Bresme2000,Vega2003}, the peculiar dependence on the temperature is specific to the one-dimensional system considered here. 
Finally, we note that the limits $N\to\infty$ and $\rho\to 1$ do not commute. Taking $\rho\to 1$ first, $-\lambda^*$ should be included in Eqs.~(\ref{eq:thermo_pf}-\ref{eq:thermo_correl}). Here, we take the thermodynamic limit at $\rho<1$ and then take the limit $\rho\to 1$.

Next, we show that the long-range charge order is necessarily zero for $\theta=0$.
For $\theta=0$ and $\theta=1/2$, the transfer matrix $\bar K_{kl}$ is even (i.e., $\bar K_{-k,-l}=\bar K_{k,l}$), so that the maximal eigenvector $\psi^*_k$ is also even. 
On the other hand, the function $k\mapsto(-1)^{\lfloor k \rfloor}$ is even for $\theta=0$ and odd for $\theta=1/2$, and so is $(\psi_-)_k$. 
Finally, the operator $\bar R$ is odd: $\bar R_{-k,-l}=-\bar R_{k,l}$.
Putting these relations together implies that $\langle\psi_-|\bar R|\psi^*\rangle=0$ for $\theta=0$: there is no long-range order and the system is in the disordered solid phase. On the other hand, long-range order can occur for $\theta=1/2$, and more generally for $\theta\neq 0$.

We have shown that the correlation length is infinite in the regions where $\theta=1/2$ for $\rho=1$. 
Here, we use an expansion around $\rho=1$ to determine its dependence on $\rho$.
We expand the operators, eigenvalues and eigenvectors as a function of the small parameter $\epsilon=1/\mu$.
The occupancy operator is $\bar R\ind{o}(\epsilon)=\bar K(\epsilon=0)$, and the density is given by
\begin{equation}\label{eq:dens_eps}
\rho(\epsilon)=\frac{\langle \psi^*(\epsilon)|\bar K(0)|\psi^*(\epsilon) \rangle}{\lambda^*(\epsilon)}=1-\epsilon \frac{{\lambda^*}'(0)}{\lambda^*(0)}+\mathcal{O}(\epsilon^2).
\end{equation}
On the other hand, the correlation length satisfies the relation
\begin{equation}\label{eq:length_eps}
\xi(\epsilon)^{-1}=\log\left(-\frac{\lambda^*(\epsilon)}{\lambda_2(\epsilon)}\right)=\epsilon \frac{{\lambda^*}'(0)+\lambda_2'(0)}{\lambda^*(0)}+\mathcal{O}(\epsilon^2),
\end{equation}
where we have used $\lambda_2(0)=-\lambda^*(0)$.
The variation of an eigenvalue, \emph{e.g.} $\lambda_2$, is given by $\lambda_2'=\langle \psi_2|\bar K'|\psi_2 \rangle$. Using that $[\psi_2(\epsilon=0)]_k=(-1)^{\lfloor k\rfloor} \psi^*(\epsilon=0)_k$ and $\bar K'(0)=\exp(-k^2/[2T])\delta(k-l)$, we get that $\lambda_2'(0)={\lambda^*}'(0)$. Inserting this equality in Eq.~(\ref{eq:length_eps}) and comparing with Eq.~(\ref{eq:dens_eps}) leads to
\begin{equation}
\xi\underset{\rho\to 1}{\sim} {\textstyle\frac{1}{2}}(1-\rho)^{-1},
\end{equation}
the correlation length is therefore half of the average distance between vacancies.

\section{Discussion}\label{}

In this Letter, we have computed exactly the bulk phase diagram of the 1d LCG (or LRPM) in the constant voltage ensemble (Fig.~\ref{fig:phase_diag}). It features a low density disordered phase and a high density ordered phase, separated by a first order phase transition at low temperatures. 
The solid state at full occupancy may be charge ordered or charge disordered, depending on the temperature.

We note that the presence of a phase transition in our one dimensional system does not violate the Landau-Peierls argument since the electrostatic interaction is long-range (or, more precisely, does not decay faster than $1/x^2$)~\cite{Landau1980,Thouless1969}).
In the transfer matrix formalism, this argument is rephrased in the van Hove theorem~\cite{Cuesta2004}. 
For the theorem to apply, the transfer matrix should be irreducible, which is not the case here because the transfer matrix can be decomposed on the subspaces $\mathcal{E}_\theta$.

Some of the bulk properties observed here may be related to the capacitor behaviour of the 1d LCG.
Firstly, the discontinuous jump of the capacitance at the point of zero charge (PZC), noticed in Ref.~\cite{Demery2012a}, which corresponds to the transition between camel- and bell-shaped capacitance curves~\cite{Kornyshev2007,Fedorov2014}, is clearly related to the first order phase transition that we observe. 
In the low density phase, integer charges are selected, so that the PZC lies in a plateau of the voltage-charge curve and the differential capacitance is small. 
In the high density phase, half-integer charges are selected and the PZC lies at the transition between two plateaus, hence the differential capacitance is large. 
Secondly, charge density oscillations were observed in the constant charge ensemble for non-integer surface charges. 
Spatial organization of the charge at an interface thus corresponds directly to charge correlation in the bulk; our results suggest that charge density oscillations seen at a charged interface are also present in the constant voltage ensemble in the regions where half-integer charges are selected in the bulk. 

Following this close relationship between the bulk and capacitor properties of the 1d lattice Coulomb gas, it would be interesting to try and extend the same approach to the three dimensional system. However, contrary to the one dimensional case, the observed layering where charge oscillates only in the direction orthogonal to the bounding surface is not directly compatible with the ordered state of the bulk liquid.
Whether the bulk properties of the RPM are related to the behaviour at interfaces thus remains as an open question.

\acknowledgments

We thank A.C. Maggs for useful discussions about the properties of the transfer matrix and the van Hove theorem. R.P. acknowledges support from the ARRS through the program P1-0055 and thanks the support of ESPCI ParisTech during his stay as a visiting \emph{Joliot Chair} Professor.


\begin{thebibliography}{10}
\expandafter\ifx\csname url\endcsname\relax\def\url#1{\texttt{#1}}\fi

\bibitem{Mezger2008}
\Name{Mezger M., Schr{\"{o}}der H., Reichert H., Schramm S., Okasinski J.~S.,
  Sch{\"{o}}der S., Honkim{\"{a}}ki V., Deutsch M., Ocko B.~M., Ralston J.,
  Rohwerder M., Stratmann M. \and Dosch H.} \REVIEW{{Science}}{322}{2008}{424}.

\bibitem{Hayes2011}
\Name{Hayes R., Borisenko N., Tam M.~K., Howlett P.~C., Endres F. \and Atkin
  R.} \REVIEW{{The Journal of Physical Chemistry C}}{115}{2011}{6855}.

\bibitem{Perkin2011}
\Name{Perkin S., Crowhurst L., Niedermeyer H., Welton T., Smith A.~M. \and
  Gosvami N.~N.} \REVIEW{{Chem. Commun.}}{47}{2011}{6572}.

\bibitem{Perkin2012}
\Name{Perkin S.} \REVIEW{{Phys. Chem. Chem. Phys.}}{14}{2012}{5052}.

\bibitem{Fedorov2014}
\Name{Fedorov M.~V. \and Kornyshev A.~A.} \REVIEW{{Chemical
  Reviews}}{114}{2014}{2978} {PMID: 24588221}.

\bibitem{Fedorov2008}
\Name{Fedorov M.~V. \and Kornyshev A.~A.} \REVIEW{{The Journal of Physical
  Chemistry B}}{112}{2008}{11868}.

\bibitem{Borukhov2000}
\Name{Borukhov I., Andelman D. \and Orland H.} \REVIEW{{Electrochemica
  acta}}{46}{2000}{221}.

\bibitem{Kornyshev2007}
\Name{Kornyshev A.~A.} \REVIEW{{Journal of Physical Chemistry
  B}}{111}{2007}{5545}.

\bibitem{Bazant2011}
\Name{Bazant M.~Z., Storey B.~D. \and Kornyshev A.~A.} \REVIEW{{Phys. Rev.
  Lett.}}{106}{2011}{046102}.

\bibitem{Demery2012}
\Name{D{\'{e}}mery V., Dean D.~S., Hammant T.~C., Horgan R.~R. \and Podgornik
  R.} \REVIEW{{EPL (Europhysics Letters)}}{97}{2012}{28004}.

\bibitem{Demery2012a}
\Name{D{\'{e}}mery V., Dean D.~S., Hammant T.~C., Horgan R.~R. \and Podgornik
  R.} \REVIEW{{The Journal of Chemical Physics}}{137}{2012}{064901}.

\bibitem{Lee2014}
\Name{Lee A.~A., Vella D., Perkin S. \and Goriely A.} \REVIEW{{The Journal of
  Chemical Physics}}{141}{2014}{}.

\bibitem{Stell1992}
\Name{Stell G.} \REVIEW{{Phys. Rev. A}}{45}{1992}{7628}.

\bibitem{Orkoulas1994}
\Name{Orkoulas G. \and Panagiotopoulos A.~Z.} \REVIEW{{The Journal of Chemical
  Physics}}{101}{1994}{1452}.

\bibitem{Vega1996}
\Name{Vega C., Bresme F. \and Abascal J. L.~F.} \REVIEW{{Phys. Rev.
  E}}{54}{1996}{2746}.

\bibitem{Stell1999}
\Name{Stell G.} \Book{{New Results on Some Ionic-Fluid Problems}} in \Book{{New
  Approaches to Problems in Liquid State Theory}}, edited by \Name{Caccamo C.,
  Hansen J.-P. \and Stell G.} Vol. 529 ({Springer Netherlands}) 1999 Ch. {NATO
  Science Series} pp. 71--89.

\bibitem{Bresme2000}
\Name{Bresme F., Vega C. \and Abascal J. L.~F.} \REVIEW{{Phys. Rev.
  Lett.}}{85}{2000}{3217}.

\bibitem{Luijten2002}
\Name{Luijten E., Fisher M.~E. \and Panagiotopoulos A.~Z.} \REVIEW{{Phys. Rev.
  Lett.}}{88}{2002}{185701}.

\bibitem{Vega2003}
\Name{Vega C., Abascal J. L.~F., McBride C. \and Bresme F.} \REVIEW{{The
  Journal of Chemical Physics}}{119}{2003}{964}.

\bibitem{Panagiotopoulos2005}
\Name{Panagiotopoulos A.~Z.} \REVIEW{{Journal of Physics: Condensed
  Matter}}{17}{2005}{}.

\bibitem{Walker1983}
\Name{Walker A.~B. \and Gillan M.~J.} \REVIEW{{Journal of Physics C: Solid
  State Physics}}{16}{1983}{3025}.

\bibitem{Panagiotopoulos1999}
\Name{Panagiotopoulos A.~Z. \and Kumar S.~K.} \REVIEW{{Phys. Rev.
  Lett.}}{83}{1999}{2981}.

\bibitem{Dickman1999}
\Name{Dickman R. \and Stell G.} \REVIEW{{AIP Conference
  Proceedings}}{492}{1999}{225}.

\bibitem{Brognara2002}
\Name{Brognara A., Parola A. \and Reatto L.} \REVIEW{{Phys. Rev.
  E}}{65}{2002}{066113}.

\bibitem{Kobelev2002}
\Name{Kobelev V., Kolomeisky A.~B. \and Fisher M.~E.} \REVIEW{{The Journal of
  Chemical Physics}}{116}{2002}{7589}.

\bibitem{Ciach2003}
\Name{Ciach A. \and Stell G.} \REVIEW{{Phys. Rev. Lett.}}{91}{2003}{060601}.

\bibitem{Diehl2005}
\Name{Diehl A. \and Panagiotopoulos A.~Z.} \REVIEW{{Phys. Rev.
  E}}{71}{2005}{046118}.

\bibitem{Ren2006}
\Name{Ren R., O{'}Keeffe C.~J. \and Orkoulas G.} \REVIEW{{The Journal of
  Chemical Physics}}{125}{2006}{}.

\bibitem{Bartsch2015}
\Name{Bartsch H., Dannenmann O. \and Bier M.} \REVIEW{{Phys. Rev.
  E}}{91}{2015}{042146}.

\bibitem{Aizenmann1981}
\Name{Aizenman M. \and Fr{\"{o}}hlich J.} \REVIEW{{Journal of Statistical
  Physics}}{26}{1981}{347} {10.1007/BF01013176}.

\bibitem{Landau1980}
\Name{Landau L.~D. \and Lifshitz E.~M.} \Book{{Statistical physics}} Vol.~5 of
  \emph{{Course of Theoretical Physics}} ({Butterworth-Heinemann}) 1980.

\bibitem{Thouless1969}
\Name{Thouless D.~J.} \REVIEW{{Phys. Rev.}}{187}{1969}{732}.

\bibitem{Cuesta2004}
\Name{Cuesta J. \and S{\'{a}}nchez A.} \REVIEW{{Journal of Statistical
  Physics}}{115}{2004}{869}.

\end{thebibliography}

\end{document}